# Title: Quantum Spin Hall Effect and Topological Field Effect Transistor in Two-Dimensional Transition Metal Dichalcogenides

**Authors:** Xiaofeng Qian[1,†], Junwei Liu[2,†], Liang Fu[2,*], and Ju Li[1,*]

**Affiliations:**

[1]Department of Nuclear Science and Engineering and Department of Materials Science and Engineering, Massachusetts Institute of Technology, Cambridge, Massachusetts 02139, USA.

[2]Department of Physics, Massachusetts Institute of Technology, Cambridge, Massachusetts 02139, USA.

*Correspondence to: liangfu@mit.edu and liju@mit.edu.

†These authors contributed equally to this work.

**Abstract**: We report a new class of large-gap quantum spin Hall insulators in two-dimensional transition metal dichalcogenides, namely, $MX_2$ with M=(Mo, W) and X=(S, Se, and Te), whose topological electronic properties are highly tunable by external electric field. We propose a novel topological field effect transistor made of these atomic layer materials and their van der Waals heterostructures. Our device exhibits parametrically enhanced charge-spin conductance through topologically protected transport channels, and can be rapidly switched off by electric field through topological phase transition instead of carrier depletion. Our work provides a practical material platform and device architecture for topological quantum electronics.

The discovery of graphene (*1*) leads to vigorous explorations of two-dimensional (2D) materials (*2*), revealing a wide range of extraordinary properties (*3-5*) and functionalities (*6, 7*). Owing to atomic thickness, 2D materials can be horizontally patterned through chemical and mechanical techniques (*8*). Moreover, the weak van der Waals (vdW) interaction between adjacent layers enables vertical stacking of different 2D materials, forming vdW heterostructures (*9*). Although being intensively explored for fundamental research and technological applications, the vast family of 2D materials and their vdW heterostructures have been largely underexploited for topological phases of matter (*10, 11*), in particular, quantum spin Hall (QSH) insulators (*12-18*).

QSH insulators have an insulating bulk but conducting edge states that are topologically protected from backscattering by time reversal symmetry. Quantized conductance through QSH edge states have been experimentally demonstrated in HgTe/CdTe (*15, 16*) and InAs/GaSb (*19, 20*) quantum wells. This could in principle provide an alternative route to quantum electronic devices with low dissipation. However, the realization of such QSH-based devices for practical applications is impeded by three critical factors: (a) band gaps of existing QSH insulators are too small, which limits the operating regime to low temperatures. This has motivated an intensive search for large-gap QSH materials (*21-25*); (b) the small number of conducting channels ($e^2/h$ per edge) results in a small signal-to-noise ratio; and (c) efficient methods of fast ON/OFF switching are lacking.

Here we show that 2D materials provide a practical platform for developing novel topological electronic devices that can potentially overcome the above hurdles. Specifically, based on first-principles calculations, we find a new class of large-gap (>0.1eV) QSH insulators in 2D transition metal dichalcogenides (TMDCs) $MX_2$ with M=(Mo, W) and X = (S, Se, Te). We demonstrate the possibility of a novel vdW-heterostructured topological field effect transistor (vdW-TFET) made of 2D atomic layer materials. Our device exhibits a parametrically enhanced conductance through QSH edge channels in the ON state, and can be rapidly switched OFF via a topological phase transition by applying an electric field. Our findings of electrically configurable QSH devices have potential applications in low-power nanoelectronics, and may enable realistic architectures of topological quantum computing based on Majorana fermions (*26-29*).

Monolayer TMDCs (*30*), $MX_2$ with M = (Mo, W) and X = (S, Se, Te), possess a variety of polytypic structures such as 2H, 1T, and 1T' (*31, 32*). These three structures are shown in Fig. 1A, 1B, and 1C, respectively. The most studied 2H structure is a sandwich of three planes of 2D hexagonally packed atoms, X-M-X, where M atoms are trigonal-prismatically coordinated by six X atoms, forming ABA stacking with the $P\bar{6}m2$ space-group. In contrast, the M atoms in the 1T structure are octahedrally coordinated with the nearby six X atoms, resulting in ABC stacking with the $P\bar{3}m1$ space group. 2H- and 1T-$MX_2$ have very different electronic properties: the former is a large gap semiconductor while the latter is metallic. It has been known that the 1T structure in $MX_2$ is typically unstable in free-standing condition and undergoes a spontaneous lattice distortion in the *x* direction to form a 2×1 superlattice structure, i.e., the 1T' structure, consisting of one-dimensional zigzag chains along the *y* direction shown in Fig. 1C (*32*). 2D

MX$_2$ in such 1T' structure, which up to now has received little attention, is the subject of this work.

The electronic structures of 1T'-MX$_2$ were obtained by first-principles calculations based on many-body perturbation theory within the GW approximation (see Methods in Supplementary Materials). Figure 2A shows the electronic band structure of 1T'-MoS$_2$, and the results of the other five compounds are shown in Supplementary Fig. S1. Unlike its 2H or 1T counterparts, 1T'-MoS$_2$ is a semiconductor with a fundamental gap of about 0.1 eV, located at $\Lambda = \pm(0, 0.146)$ Å$^{-1}$ and indicated by red dots in Fig. 2B. The conduction and valence bands display a camelback shape near $\Gamma$ in the 2D Brillouin zone (BZ, see Fig. 2B), suggestive of band inversion with a large inverted gap at $\Gamma$ over 0.6 eV. Since the 1T' structure has inversion symmetry, the Z$_2$ band topology can be determined by the parity of valence bands at four time-reversal invariant momenta (TRIM), $\Gamma$, X, Y and R (33). We calculated these band parity indices for 1T'-MoS$_2$ (Fig. 2C) and found that the product of parity is -1. This demonstrates the Z$_2$ nontrivial band topology and establishes the QSH insulator phase in 2D 1T'-MX$_2$.

To understand the nature of the inverted bands near $\Gamma$, we examined their orbital characters, and found the valence band mainly consists of $d$-orbitals of M atoms ($d_{yz}$ and $d_{xy}$), while the conduction band mainly consists of $p_y$-orbitals of X atoms. By analyzing the symmetry properties of these bands, we develop a low-energy $\mathbf{k} \cdot \mathbf{p}$ Hamiltonian for 1T'-MX$_2$,

$$H = \begin{pmatrix} E_p(k_x, k_y) & 0 & -iv_1\hbar k_x & v_2\hbar k_y \\ 0 & E_p(k_x, k_y) & v_2\hbar k_y & -iv_1\hbar k_x \\ iv_1\hbar k_x & v_2\hbar k_y & E_d(k_x, k_y) & 0 \\ v_2\hbar k_y & iv_1\hbar k_x & 0 & E_d(k_x, k_y) \end{pmatrix},$$

where $E_p = -\delta - \frac{\hbar^2 k_x^2}{2m_x^p} - \frac{\hbar^2 k_y^2}{2m_y^p}$, and $E_d = \delta + \frac{\hbar^2 k_x^2}{2m_x^d} + \frac{\hbar^2 k_y^2}{2m_y^d}$. Here $\delta < 0$ corresponds to the $d$-$p$ band inversion. By fitting with first-principles band structure in Fig. 2A, we obtain the parameters: $\delta = -0.33$ eV, $v_1 = 3.87 \times 10^5$ m/s, $v_2 = 0.46 \times 10^5$ m/s, $m_x^p = 0.50\, m_e$, $m_y^p = 0.16\, m_e$, $m_x^d = 2.48\, m_e$, and $m_y^d = 0.37\, m_e$, where $m_e$ is the free electron mass.

In addition to MoS$_2$, we found that all other five MX$_2$ in the 1T' structure: MoSe$_2$, MoTe$_2$, WS$_2$, WSe$_2$ and WTe$_2$, have Z$_2$ nontrivial band topology resulting from the above $p$-$d$ band inversion (see Fig. S2 of Supplementary Materials), with inverted band gaps at $\Gamma$ of 1.04, 0.36, 0.28, 0.94, and 1.17 eV, respectively. 1T'-MoSe$_2$, WS$_2$, WSe$_2$ have fundamental gaps of 0.11, 0.12, and 0.12 eV, respectively. On the other hand, 1T'-MoTe$_2$ and WTe$_2$ are semi-metals due to the increase of valence band maximum at the $\Gamma$ point (see Fig. S1 of Supplementary Materials).

The QSH insulator phase in 2D 1T'-MX$_2$ leads to helical edge states that are protected from elastic backscattering by TRS. To demonstrate these edge states explicitly, we carried out iterative Green's function calculations using first-principles tight binding Hamiltonian (*34, 35*) constructed from many-body GW theory (see Methods in Supplementary Materials). Figure 2D shows the edge states of monolayer 1T'-MoS$_2$, where a Dirac point at Γ is located inside the band gap with a high velocity of ~ $1.0\times10^5$ m/s. The counter-propagating edge modes exhibit opposite spin polarizations, in accordance with the spin-momentum locking of one-dimensional helical electrons. From the local density of states as a function of distance away from the edge (Fig. 2E), we find the decay length of helical edge states around Γ is about 5 nm, which is much shorter than the case of HgTe quantum wells (50 nm) (*36*). Due to such highly localized nature, helical edge states in 1T'-MX$_2$ will experience significantly reduced scattering with bulk states (*37*), and hence have increased transport lifetime. Similar results are found for five other TMDCs, as shown in the Fig. S3 and S4 of Supplementary Materials.

From the perspective of potential device applications, the ability to control topological electronic properties is highly desirable. Based on first-principles calculations, we investigated the effect of a vertical electric field on the electronic structure of 1T'-MoS$_2$ QSH insulator. Figure 3A displays bulk band structures along Y-Γ-Y under different electric fields from 0 to 0.2 V/Å, while Figure 3B shows the corresponding edge states. The electric field breaks the inversion symmetry and introduces Rashba spin splitting of the original doubly degenerate bands near the fundamental gap at Λ. As the field increases, the band gap of monolayer 1T'-MoS$_2$ first decreases, closes at a critical field of 0.142 V/Å, and then reopens (see Fig. 4A). This induces a topological phase transition to a trivial insulator via Dirac fermion mass reversal (*38*) and thus leads to the destruction of helical edge states, shown in Fig. 3B. This is confirmed by our numerical calculation of the Z$_2$ invariant (*39*). In addition to the electrical field, we also find a few percent of in-plane elastic strain can change 1T'-MoTe$_2$ and WTe$_2$ from semi-metals to small-gap QSH insulators by lifting the band overlap. Since monolayer 1T'-MX$_2$ can sustain large elastic strains up to ~10%, mechanical strain provides an alternative route to Dirac band gap engineering, in addition to the electric field.

The above findings of field-induced electronic transition and topology change enables all-electrical control of the ON/OFF charge/spin conductance of helical edge states, which have significant implications on the design of QSH-based transistors and other novel devices (*40*). To illustrate the feasibility of device applications, we propose a topological field effect transistor (TFET) based on van der Waals heterostructures of 2D TMDCs and 2D wide-gap insulators based on the detailed phase diagram shown in Fig. 4A. The schematic device is shown in Fig. 4B, where the top and bottom gate supply the vertical electric field to control the ON/OFF function. For practical applications, it is advantageous to use multilayers of 1T'-MX$_2$ QSH insulators, which increases the number of edge transport channels proportional to the number of layers. Adjacent QSH layers are separated by 2D wide-gap insulators such hexagonal boron nitride (*h*BN) (*41*). The insulating *h*BN layers serve to electrically decouple different QSH layers and protect parallel helical edge channels from being gapped by inter-layer hybridization. Under ideal conditions, this device will support dissipationless charge/spin transport in the ON state (Z$_2$=1) with a quantized conductance of $2Ne^2/h$, where *N* is the number of QSH layers. Applying a moderate electric field will transform 1T'-MX$_2$ into an ordinary insulator (Z$_2$=0) and turn the edge conduction OFF, see Fig. 4A.

The proposed vdW-TFET may possess important technical advantages, due to its novel operation mechanism that is fundamentally different from traditional metal-oxide-semiconductor field-effect transistors (MOS-FET). MOS-FET works by injecting/depleting free carrier beneath the gate, with a *RC*-type response time influenced by carrier concentration and mobility in the working channel, while our vdW-TFET works by switching the presence/absence of topologically protected conduction channels without changing the carrier density. The electric field induced topological phase transition can happen very rapidly, with electronic response timescale (*42, 43*). In addition, the vdW heterostructure and the 2D nature of vdW-TFET make it convenient to *vertically* and *horizontally* pattern devices in large scale. As the decay length of edge states in monolayer 1T'-$MoS_2$ is only 5 nm, the minimum width of vdW-TFET is around 10 nm. TFET devices of such nanoscale size have large surface-to-bulk ratio, which will greatly reduce the contribution of thermally populated bulk carriers to the total electric current and hence enhance the ON/OFF ratio.

Encouragingly, from materials perspective, the desired 1T' structure in monolayer $MoS_2$ has recently been observed in high-resolution TEM experiments (*32*). Also, 1T' is known to be the underlying ground-state structure of layered $WTe_2$ (*30*). To verify the structural stability of 1T' in these 2D TMDCs, we carried out first-principles calculations of energy barrier from 2H to 1T' and from 1T' to 1T. Figure 5A shows the calculated relative total energy per $MX_2$ compared to their 2H structures. We found monolayer $WTe_2$ is more stable in 1T' structure, while the other five $MX_2$, including $MoS_2$, $MoSe_2$, $MoTe_2$, and $WS_2$, $WSe_2$, are more stable in the commonly found 2H structure. Nonetheless, there exists a large energy barrier of more than 1eV/$MX_2$ between 1T' and 2H for all $MX_2$, suggesting that the 1T' structure can be stabilized under appropriate chemical, thermal, or mechanical conditions. Moreover, we found the 1T structure is highly unstable compared to 1T', with an energy difference from 0.35 eV/$MoS_2$ up to 0.7 eV/$WTe_2$. Our calculation shows no energy barrier from 1T to 1T' structure, implying a spontaneous structural relaxation of 1T to 1T' in the monolayer $MX_2$. In addition, we verified the stability of the 1T' structure by computing its phonon dispersion, shown in Fig. 5B for 1T'-$MoS_2$ and Fig. S5 for other 2D TMDCs. The absence of imaginary frequency throughout the 2D BZ provides a direct evidence of the structural stability in all 1T'-$MX_2$.

Because of its extraordinarily large inverted gap of ~0.6 eV, the QSH insulator phase in 2D TMDCs is robust against structural deformations and imperfections such as long-wave length wrinkles and ripples. Unlike HgTe/CdTe and InAs/GaSb quantum wells that require exquisite growth by molecular beam epitaxy, stable 2D TMDCs can be mechanically exfoliated from bulk materials and chemically synthesized as well. Various electronic devices based on atomic layer TMDCs and their vdW heterostructures have been recently demonstrated. Thus our discovery of topological phases of matter in 2D TMDCs opens up new directions for research and development in both fields.

**References and Notes:**

1. K. S. Novoselov, A. K. Geim, S. V. Morozov, D. Jiang, Y. Zhang, S. V. Dubonos, I. V. Grigorieva, A. A. Firsov, Electric Field Effect in Atomically Thin Carbon Films. *Science* **306**, 666-669 (2004).
2. K. S. Novoselov, D. Jiang, F. Schedin, T. J. Booth, V. V. Khotkevich, S. V. Morozov, A. K. Geim, Two-dimensional atomic crystals. *Proc. Natl. Acad. Sci. U. S. A.* **102**, 10451-10453 (2005).
3. C. Lee, X. Wei, J. W. Kysar, J. Hone, Measurement of the Elastic Properties and Intrinsic Strength of Monolayer Graphene. *Science* **321**, 385-388 (2008).
4. A. H. Castro Neto, F. Guinea, N. M. R. Peres, K. S. Novoselov, A. K. Geim, The electronic properties of graphene. *Rev. Mod. Phys.* **81**, 109-162 (2009).
5. K. F. Mak, C. Lee, J. Hone, J. Shan, T. F. Heinz, Atomically Thin $MoS_2$: A New Direct-Gap Semiconductor. *Phys. Rev. Lett.* **105**, 136805 (2010).
6. Q. H. Wang, K. Kalantar-Zadeh, A. Kis, J. N. Coleman, M. S. Strano, Electronics and optoelectronics of two-dimensional transition metal dichalcogenides. *Nature Nanotech.* **7**, 699-712 (2012).
7. J. Feng, X. Qian, C.-W. Huang, J. Li, Strain-engineered artificial atom as a broad-spectrum solar energy funnel. *Nature Photon.* **6**, 866-872 (2012).
8. K. S. Kim, Y. Zhao, H. Jang, S. Y. Lee, J. M. Kim, K. S. Kim, J.-H. Ahn, P. Kim, J.-Y. Choi, B. H. Hong, Large-scale pattern growth of graphene films for stretchable transparent electrodes. *Nature* **457**, 706-710 (2009).
9. A. K. Geim, I. V. Grigorieva, Van der Waals heterostructures. *Nature* **499**, 419-425 (2013).
10. D. Kong, Y. Cui, Opportunities in chemistry and materials science for topological insulators and their nanostructures. *Nature Chemistry* **3**, 845-849 (2011).
11. K. Yang, W. Setyawan, S. Wang, M. Buongiorno Nardelli, S. Curtarolo, A search model for topological insulators with high-throughput robustness descriptors. *Nature Mater.* **11**, 614-619 (2012).
12. C. L. Kane, E. J. Mele, Quantum Spin Hall Effect in Graphene. *Phys. Rev. Lett.* **95**, 226801 (2005).
13. C. L. Kane, E. J. Mele, $Z_2$ Topological Order and the Quantum Spin Hall Effect. *Phys. Rev. Lett.* **95**, 146802 (2005).
14. B. A. Bernevig, S.-C. Zhang, Quantum Spin Hall Effect. *Phys. Rev. Lett.* **96**, 106802 (2006).
15. B. A. Bernevig, T. L. Hughes, S.-C. Zhang, Quantum Spin Hall Effect and Topological Phase Transition in HgTe Quantum Wells. *Science* **314**, 1757-1761 (2006).
16. M. König, S. Wiedmann, C. Brüne, A. Roth, H. Buhmann, L. W. Molenkamp, X.-L. Qi, S.-C. Zhang, Quantum Spin Hall Insulator State in HgTe Quantum Wells. *Science* **318**, 766-770 (2007).
17. M. Z. Hasan, C. L. Kane, Topological insulators. *Rev. Mod. Phys.* **82**, 3045-3067 (2010).
18. X.-L. Qi, S.-C. Zhang, Topological insulators and superconductors. *Rev. Mod. Phys.* **83**, 1057-1110 (2011).
19. L. Du, I. Knez, G. Sullivan, R.-R. Du, Observation of Quantum Spin Hall States in InAs/GaSb Bilayers under Broken Time-Reversal Symmetry. http://arxiv.org/abs/1306.1925, (2013).
20. C. Liu, T. L. Hughes, X.-L. Qi, K. Wang, S.-C. Zhang, Quantum Spin Hall Effect in Inverted Type-II Semiconductors. *Phys. Rev. Lett.* **100**, 236601 (2008).

21. S. Murakami, Quantum Spin Hall Effect and Enhanced Magnetic Response by Spin-Orbit Coupling. *Phys. Rev. Lett.* **97**, 236805 (2006).
22. C.-C. Liu, W. Feng, Y. Yao, Quantum Spin Hall Effect in Silicene and Two-Dimensional Germanium. *Phys. Rev. Lett.* **107**, 076802 (2011).
23. D. Xiao, W. Zhu, Y. Ran, N. Nagaosa, S. Okamoto, Interface engineering of quantum Hall effects in digital transition metal oxide heterostructures. *Nature Commun.* **2**, 596 (2011).
24. Y. Xu, B. Yan, H.-J. Zhang, J. Wang, G. Xu, P. Tang, W. Duan, S.-C. Zhang, Large-Gap Quantum Spin Hall Insulators in Tin Films. *Phys. Rev. Lett.* **111**, 136804 (2013).
25. H. Weng, X. Dai, Z. Fang, Transition-Metal Pentatelluride $ZrTe_5$ and $HfTe_5$: A Paradigm for Large-Gap Quantum Spin Hall Insulators. *Physical Review X* **4**, 011002 (2014).
26. L. Fu, C. L. Kane, Superconducting Proximity Effect and Majorana Fermions at the Surface of a Topological Insulator. *Phys. Rev. Lett.* **100**, 096407 (2008).
27. L. Fu, C. L. Kane, Josephson current and noise at a superconductor/quantum-spin-Hall-insulator/superconductor junction. *Phys. Rev. B* **79**, 161408 (2009).
28. S. Mi, D. I. Pikulin, M. Wimmer, C. W. J. Beenakker, Proposal for the detection and braiding of Majorana fermions in a quantum spin Hall insulator. *Phys. Rev. B* **87**, 241405 (2013).
29. J. Alicea, New directions in the pursuit of Majorana fermions in solid state systems. *Rep. Prog. Phys.* **75**, 076501 (2012).
30. J. A. Wilson, A. D. Yoffe, The transition metal dichalcogenides discussion and interpretation of the observed optical, electrical and structural properties. *Advances in Physics* **18**, 193-335 (1969).
31. J. Heising, M. G. Kanatzidis, Exfoliated and Restacked $MoS_2$ and $WS_2$: Ionic or Neutral Species? Encapsulation and Ordering of Hard Electropositive Cations. *J. Am. Chem. Soc.* **121**, 11720-11732 (1999).
32. G. Eda, T. Fujita, H. Yamaguchi, D. Voiry, M. Chen, M. Chhowalla, Coherent Atomic and Electronic Heterostructures of Single-Layer $MoS_2$. *ACS Nano* **6**, 7311-7317 (2012).
33. L. Fu, C. L. Kane, Topological insulators with inversion symmetry. *Phys. Rev. B* **76**, 045302 (2007).
34. N. Marzari, D. Vanderbilt, Maximally localized generalized Wannier functions for composite energy bands. *Phys. Rev. B* **56**, 12847-12865 (1997).
35. X. Qian, J. Li, L. Qi, C.-Z. Wang, T.-L. Chan, Y.-X. Yao, K.-M. Ho, S. Yip, Quasiatomic orbitals for ab initio tight-binding analysis. *Phys. Rev. B* **78**, 245112 (2008).
36. B. Zhou, H.-Z. Lu, R.-L. Chu, S.-Q. Shen, Q. Niu, Finite Size Effects on Helical Edge States in a Quantum Spin-Hall System. *Phys. Rev. Lett.* **101**, 246807 (2008).
37. J. I. Väyrynen, M. Goldstein, L. I. Glazman, Helical Edge Resistance Introduced by Charge Puddles. *Phys. Rev. Lett.* **110**, 216402 (2013).
38. S. Murakami, Phase transition between the quantum spin Hall and insulator phases in 3D: emergence of a topological gapless phase. *New J. Phys.* **9**, 356-356 (2007).
39. T. Fukui, Y. Hatsugai, Quantum Spin Hall Effect in Three Dimensional Materials: Lattice Computation of $Z_2$ Topological Invariants and Its Application to Bi and Sb. *J. Phys. Soc. Jpn.* **76**, 053702 (2007).
40. J. Wunderlich, B.-G. Park, A. C. Irvine, L. P. Zârbo, E. Rozkotová, P. Nemec, V. Novák, J. Sinova, T. Jungwirth, Spin Hall Effect Transistor. *Science* **330**, 1801-1804 (2010).
41. C. R. Dean, A. F. Young, I. Meric, C. Lee, L. Wang, S. Sorgenfrei, K. Watanabe, T. Taniguchi, P. Kim, K. L. Shepard, J. Hone, Boron nitride substrates for high-quality graphene electronics. *Nature Nanotech.* **5**, 722-726 (2010).

42. J. Liu, T. H. Hsieh, P. Wei, W. Duan, J. Moodera, L. Fu, Spin-filtered edge states with an electrically tunable gap in a two-dimensional topological crystalline insulator. *Nature Mater.* **13**, 178-183 (2014).
43. P. Michetti, B. Trauzettel, Devices with electrically tunable topological insulating phases. *Appl. Phys. Lett.* **102**, 063503 (2013).

**Acknowledgments:** We acknowledge support from NSF under Award DMR-1120901 (X.Q. and J.L.), and U.S. Department of Energy, Office of Basic Energy Sciences, Division of Materials Sciences and Engineering under Award DE-SC0010526 (L.F. and J.W.L). Computational time on the Extreme Science and Engineering Discovery Environment (XSEDE) under the grant number TG-DMR130038 and TG-DMR140003 is gratefully acknowledged. X.Q. and J.W.L. contributed equally to this work.

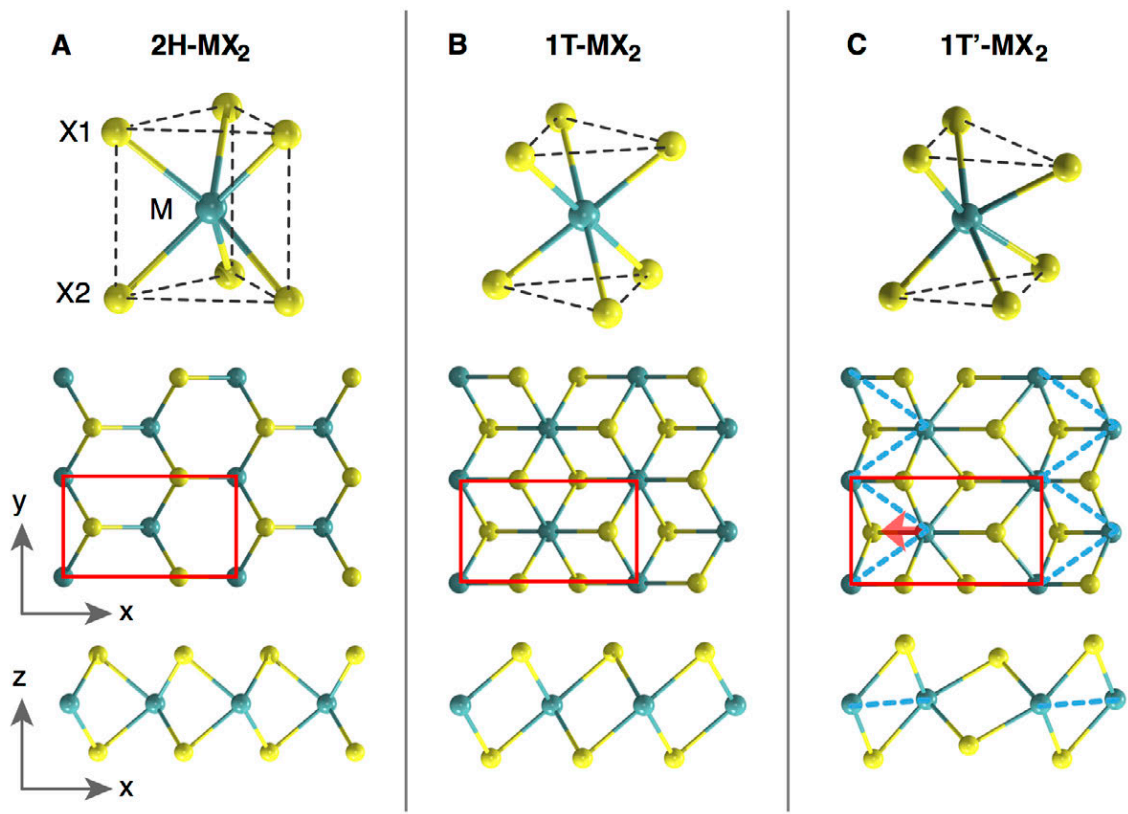

**Fig. 1**. Atomistic structure of monolayer transition metal dichalcogenide $MX_2$. M stands for (Mo,W) and X stands for (S, Se, Te). (**A**) 2H-$MX_2$ where M atoms are trigonal-prismatically coordinated by six X atoms, forming ABA stacking with the $P\bar{6}m2$ space-group. (**B**) 1T-$MX_2$ where M atoms are octahedrally coordinated with the nearby six X atoms, forming ABC stacking with the $P\bar{3}m1$ space group. (**C**) 1T'-$MX_2$, distorted 1T-$MX_2$, where the distorted M atoms form one-dimensional zigzag chains indicated by the dashed blue line. Their unit cell is indicated by red rectangles.

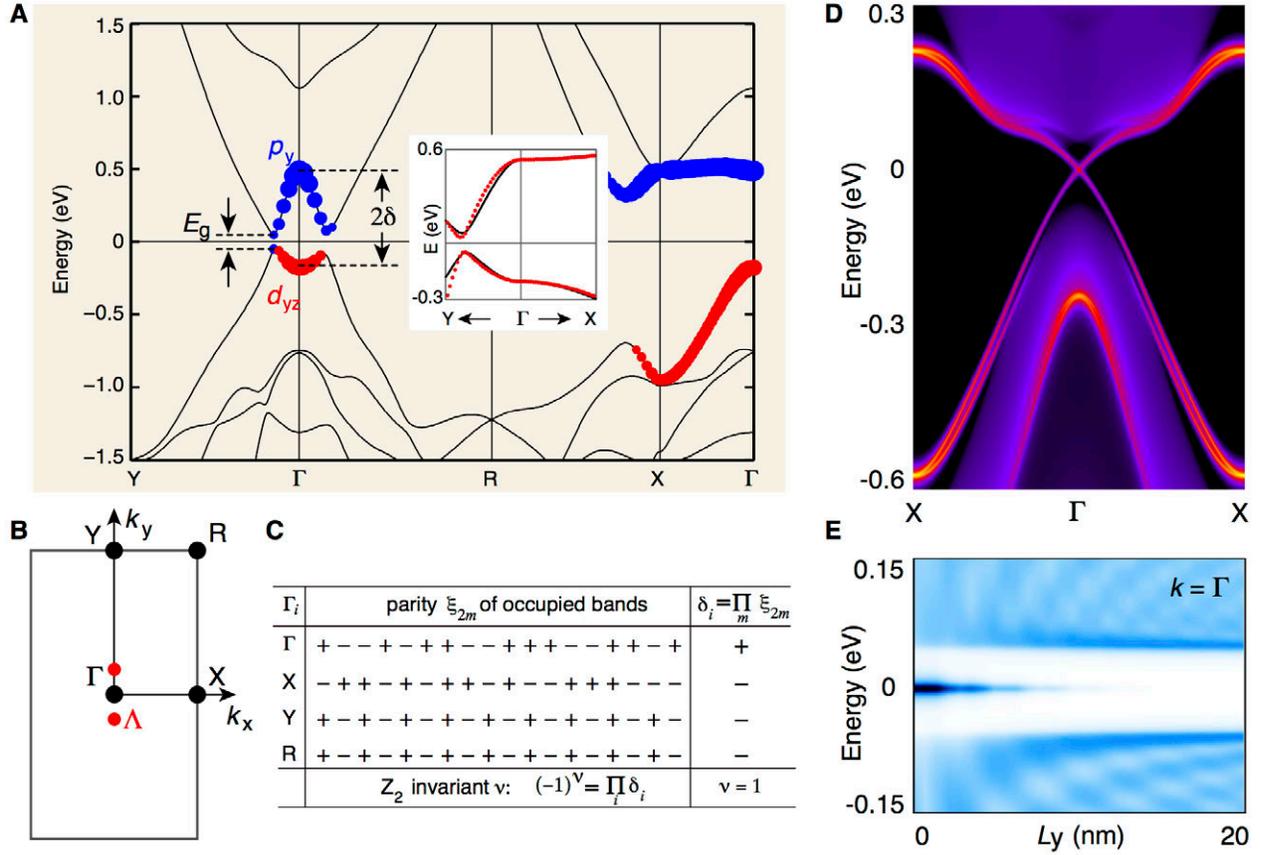

**Fig. 2.** Electronic structures of monolayer 1T'-MoS$_2$. (**A**) Electronic band structure along high symmetry lines of Brillouin zone. $E_g$ and $2\delta$ represent the fundamental gap and inverted gap, respectively. Blue and red dots indicate the major orbital characters in the top valence band and bottom conduction band. The inset compares the band structures from first-principles calculation (black line) and the fitted $k \cdot p$ Hamiltonian (red dots). (**B**) Brillouin zone of 1T'-MoS$_2$. Four time-reversal invariant momenta are marked by black dots and labeled as Γ, X, Y, and R. The locations of fundamental gap are marked by red dots and labeled by Λ. (**C**) Parity table of electronic bands at time-reversal invariant momenta: Γ, X, Y, and R. (**D**) Edge density of states and (**E**) local density of states at Γ point as a function of penetration depth $L_y$ away from edge where a clear sharp peak from the edge states appears in the gap with a decay length of ~5 nm.

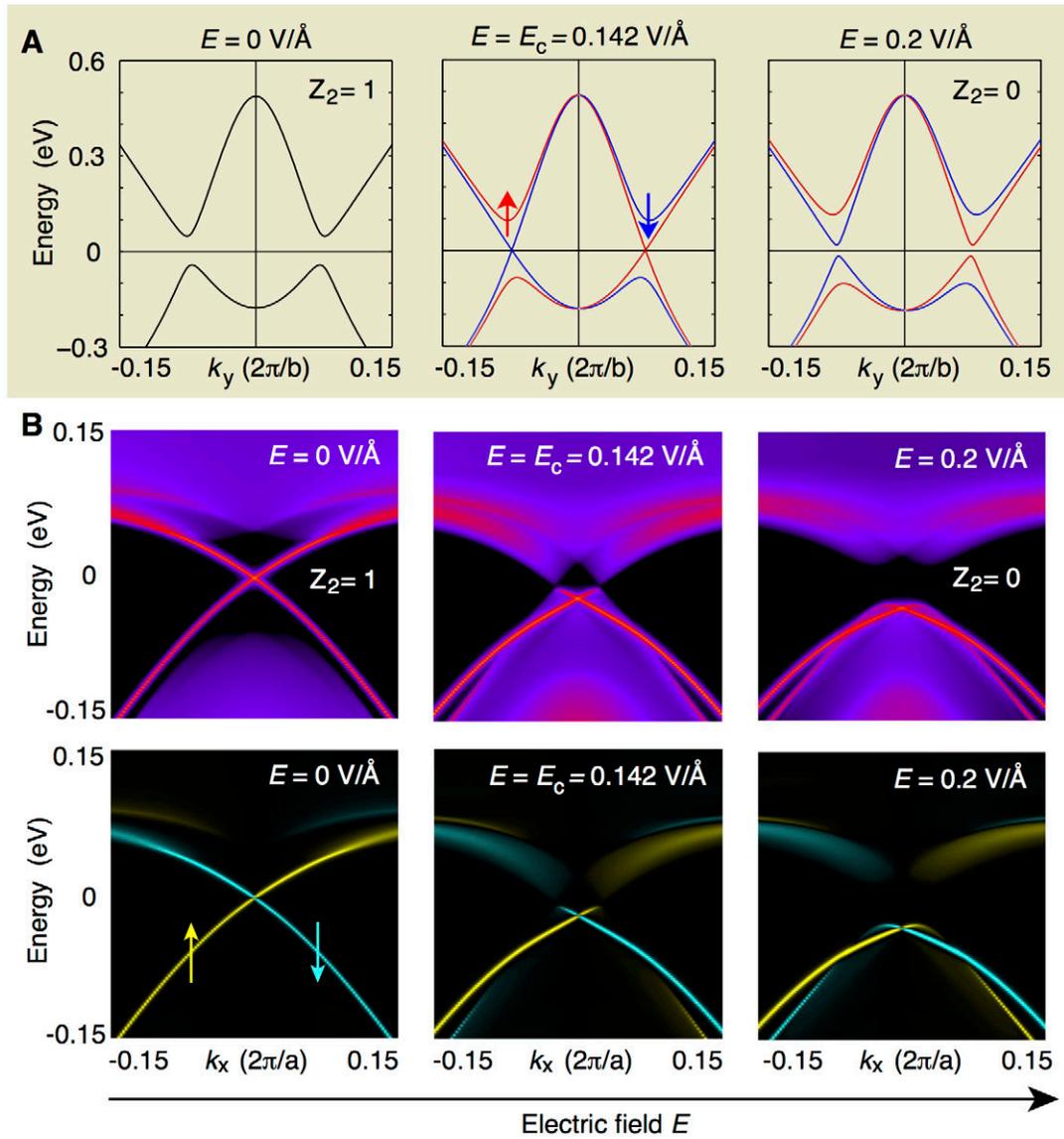

**Fig. 3.** Vertical electric field induced topological phase transition in monolayer 1T'-MoS$_2$. (**A**) Bulk band structure of monolayer 1T'-MoS$_2$ under different electric field. (**B**) Edge density of states (second row) and spin polarization (third row) under different electric field.

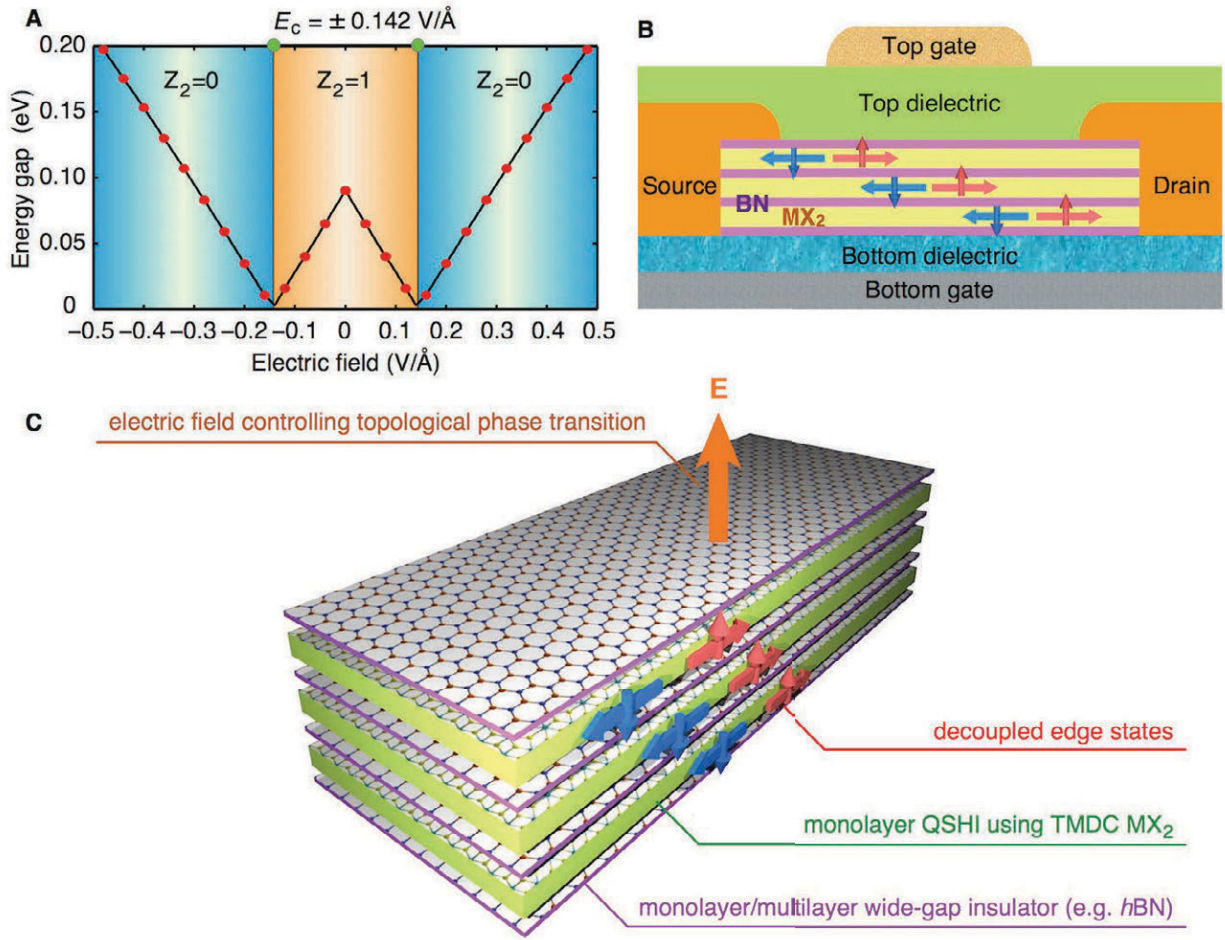

**Fig. 4.** Van der Waals heterostructured topological field effect transistor. (**A**) Topological phase diagram of monolayer 1T'-MoS$_2$ as function of vertical electric field. The critical field strength is ±0.142 V/Å, marked by two green dots. (**B**) Schematic of van der Waals heterostructured field effect transistor. The central component is the van der Waals heterostructure of alternating monolayer transition metal dichalcogenide 1T'-MX$_2$ and mono-/multi-layer wide-gap insulators such as $h$BN. Carriers are injected from source electrode and ejected from drain electrode. Depending on specific application, the carriers can be charge or spin. The ON/OFF switch is controlled by vertical electric field through the top and bottom gates. They are separated with the central van der Waals heterostructure by top and bottom dielectrics to minimize the leakage current. (**C**) Schematic of working mechanism. The mono-/multi-layer wide-gap insulators such as $h$BN can effectively screen the interaction between adjacent layers of transition metal dichalcogenide MX$_2$, preventing them from detrimental topological phase change. This vdW heterostructure can parametrically increase the number of edge channels and thus significantly enhance the signal-to-noise ratio in practical applications.

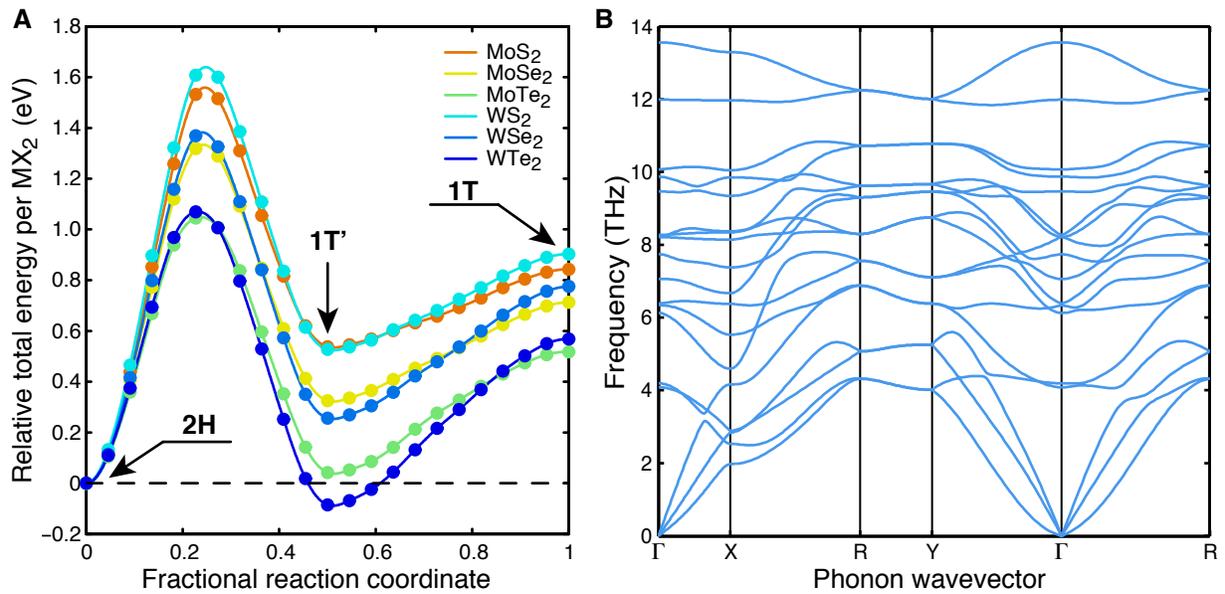

**Fig. 5.** Structural stability of monolayer transition metal dichalcogenide 1T'-$MX_2$. (**A**) Relative total energy per $MX_2$ as a function of fractional reaction coordinate. (**B**) Phonon band structure of monolayer 1T'-$MoS_2$. The absence of imaginary frequency indicates the stability of monolayer 1T'-$MoS_2$.

**Supplementary Materials:**

Methods

Figures S1-S5

References

# Supporting Online Material for

**Quantum Spin Hall Effect and Topological Field Effect Transistor in Two-Dimensional Transition Metal Dichalcogenides**

Xiaofeng Qian[1,†], Junwei Liu[2,†], Liang Fu[2,*], and Ju Li[1,*]

[1]Department of Nuclear Science and Engineering and Department of Materials Science and Engineering, Massachusetts Institute of Technology, Cambridge, Massachusetts 02139, USA.

[2]Department of Physics, Massachusetts Institute of Technology, Cambridge, Massachusetts 02139, USA.

*Correspondence to: liangfu@mit.edu and liju@mit.edu.

†These authors contributed equally to this work.

## Methods:

### 1. Ground-state atomic structures of monolayer transition metal dichalcogenides 1T'-MX$_2$

Ground-state atomic structures of all six monolayer transition metal dichalcogenides 1T'-MX$_2$ were fully relaxed using first-principles density functional theory (DFT) (*1, 2*). The calculations were performed by the Vienna Ab initio Simulation Package (*3, 4*) with projector-augmented wave method (*5*) and exchange-correlation functional in the Perdew-Berke-Ernzerhof's form (*6*) within the generalized-gradient approximation (*7, 8*). We used an energy cutoff of 400 eV and maximum residual force less than 0.001 eV/Å. The Monkhorst-Pack *k*-point sampling (*9*) of 8×8×1 was adopted for the Brillouin zone integration, and a large vacuum region of more than 16 Å was applied to the plane normal direction in order to minimize image interactions from the periodic boundary condition. Moreover, phonon dispersion curves were calculated by first-principles density-functional perturbation theory (*10*) and the results are shown in Fig. 5B for monolayer 1T'-MoS$_2$ and Fig. S5 for all six materials.

### 2. Bulk electronic structure of monolayer transition metal dichalcogenides 1T'-MX$_2$

We first computed the electronic band structures of monolayer transition metal dichalcogenides 1T'-MX$_2$ using first-principles DFT. Since the calculated DFT band structures only provide qualitative electronic structure for quasi-particles such as electrons and holes, we further computed quasi-particle band structures using more accurate method, namely, many-body perturbation theory within Hedin's G$_0$W$_0$ approximation (*11, 12*). The results are shown in Fig. 2A for monolayer 1T'-MoS$_2$ and Fig. S1 for all six materials.

### 3. Electronic structure of edge states in monolayer transition metal dichalcogenides 1T'-MX$_2$

The Z$_2$ trivial/nontrivial band topology has distinct consequence on the helical edge state. To reveal the helical edge states of monolayer 1T'-MX$_2$ explicitly, we performed iterative Green's function calculations (*13*) using tight binding Hamiltonian (*14, 15*) constructed from many-body perturbation theory with the GW approximation (*11, 12*), where we extracted the edge density of states, spin polarization, and decay length of the helical edge states. The results are shown in Fig. 3 for monolayer 1T'-MoS$_2$ and Fig. S3 and Fig. S4 for all six materials. Furthermore, to investigate the effect of vertical electric field on the electronic structure of monolayer 1T'-MoS$_2$ QSHI, we introduce a corresponding change in the diagonal elements of first-principles tight binding Hamiltonian. This approach is validated by comparing with direct self-consistent first-principles calculations under the same electric field. In addition, we found the electric field has negligible impact on ionic positions.

### 4. Z$_2$ invariant of monolayer transition metal dichalcogenides 1T'-MX$_2$

The Z$_2$ invariant was obtained by explicitly calculating band parity of the materials with inversion symmetry (*16*). The results are shown in Fig. 2C for monolayer 1T'-MoS$_2$ and Fig. S2 for all six materials. We also cross-checked all the results by the *n*-field method (*17*).

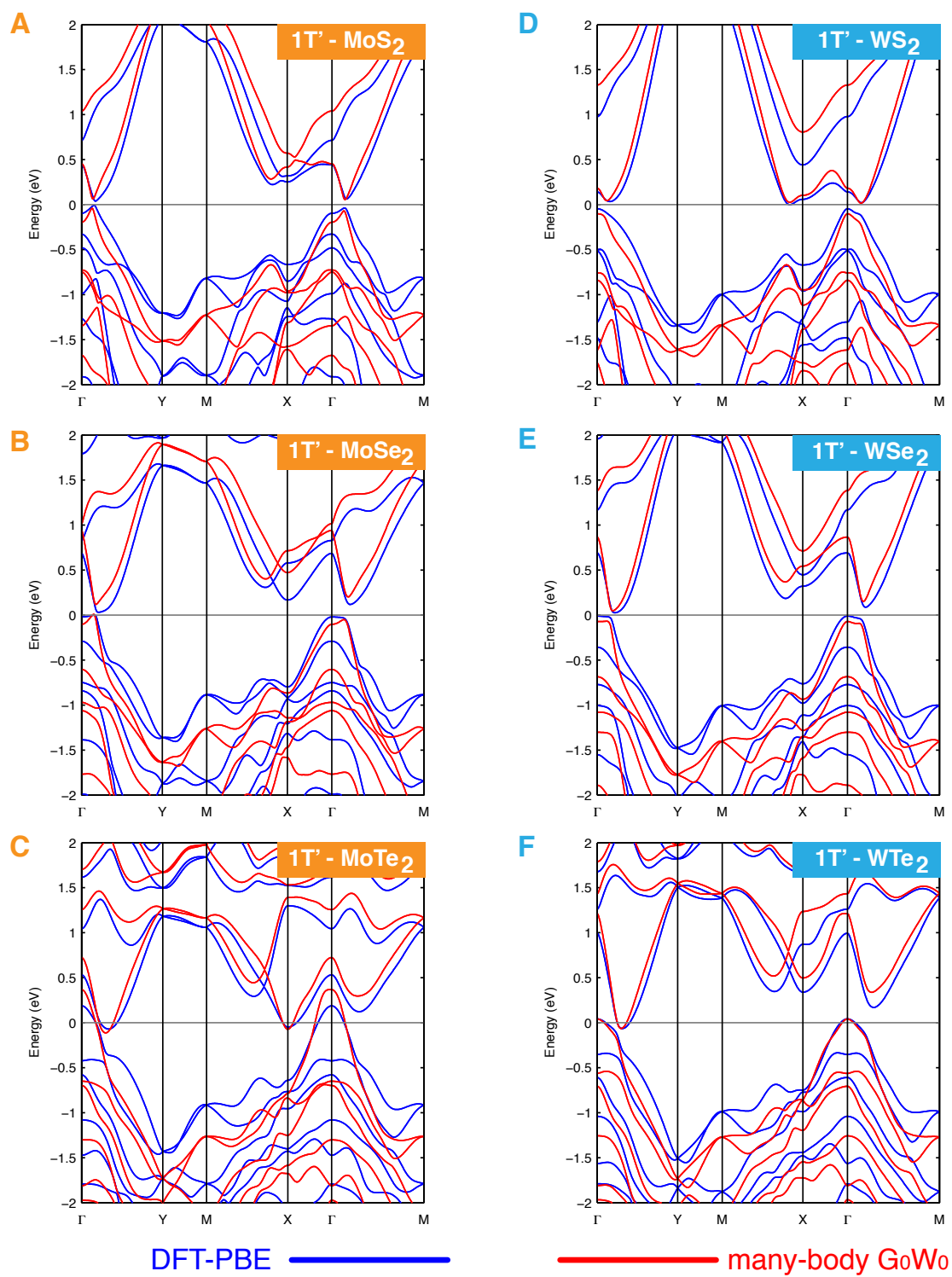

Fig. S1. Electronic band structure of monolayer transition metal dichalcogenides 1T'-$MX_2$. Blue lines stand for first-principles density-functional theory calculations. Red lines stand for many-body $G_0W_0$ theory calculations.

Fig. S2. Parity tables for monolayer transition metal dichalcogenides 1T'-MX$_2$ at time-reversal invariant momenta, and the corresponding Z$_2$ invariants, $\nu$. All Z$_2$ invariants are 1, indicating all six 1T'-MX$_2$ are topologically nontrivial.

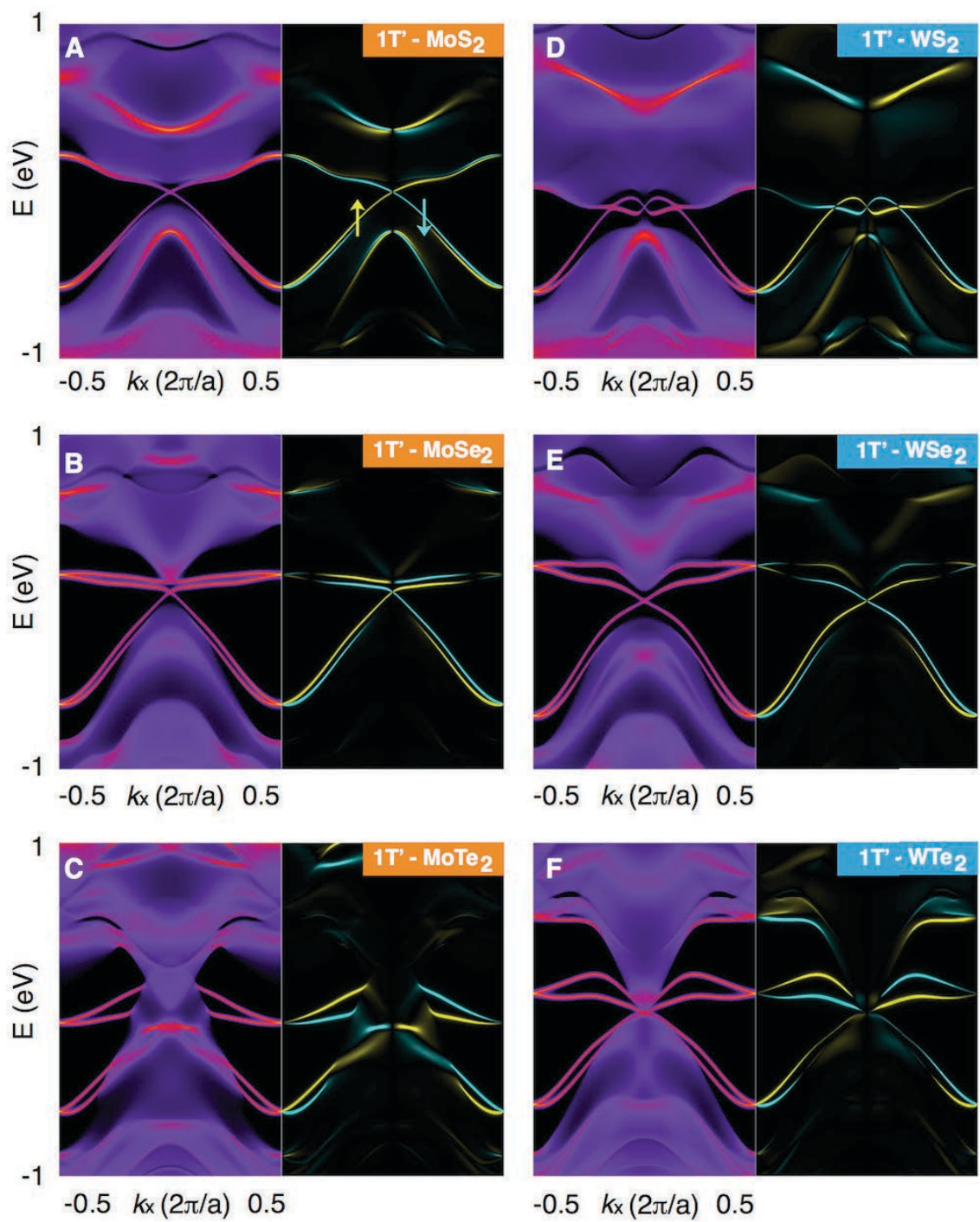

Fig. S3. Electronic structure of helical edge states in monolayer transition metal dichalcogenides 1T'-MX$_2$. For each MX$_2$, the left subpanel shows the density of states while the right subpanel shows the corresponding spin polarization. These helical edge states are a manifestation of nontrivial topology of monolayer 1T'-MX$_2$.

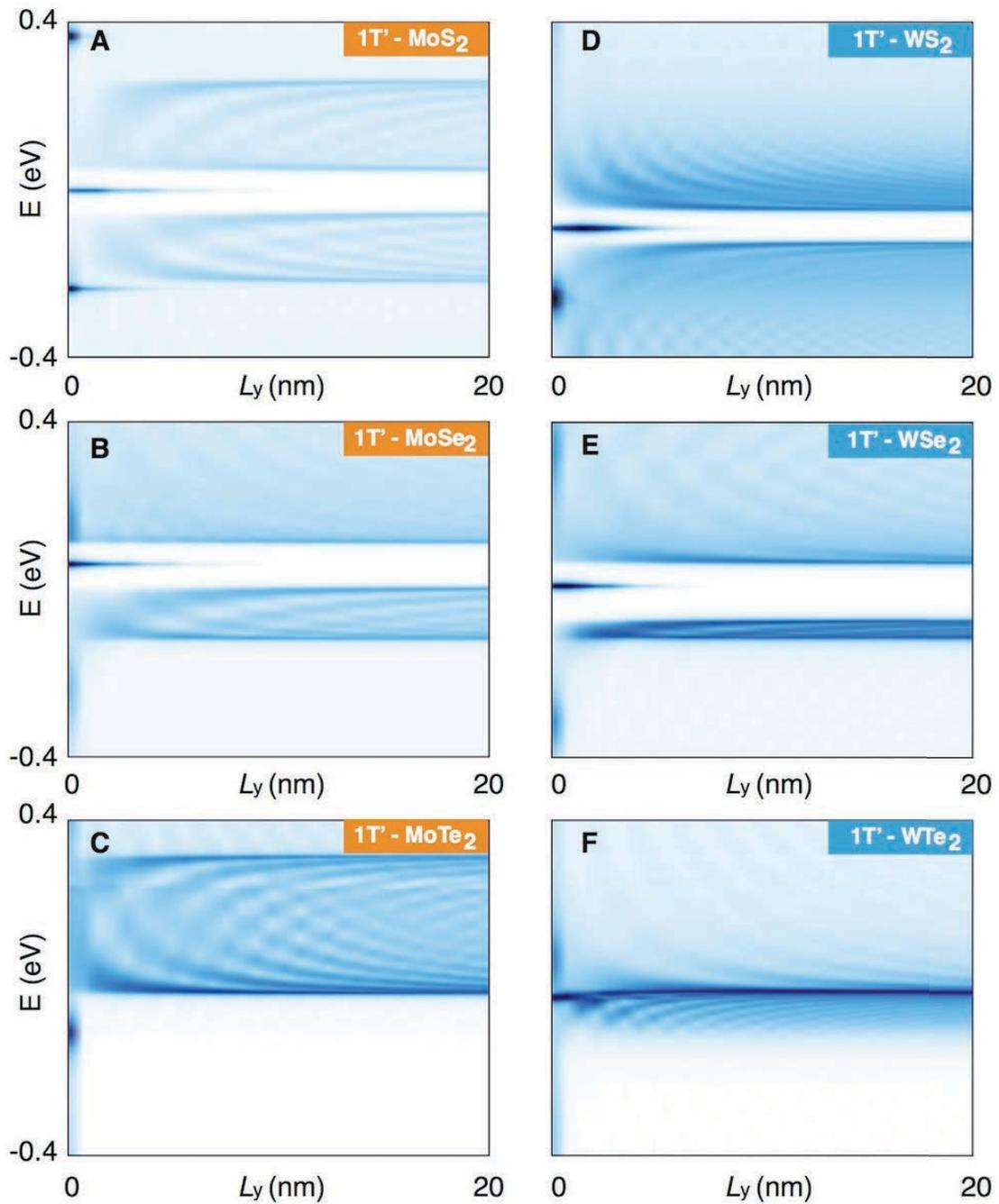

Fig. S4. Local density of states as function of penetration depth away from the edge at $L_y=0$ in monolayer transition metal dichalcogenide 1T'-$MX_2$. Due to the semi-metallic nature of 1T'-$MoTe_2$ and 1T'-$WTe_2$, local density of states from edge and bulk are entangled with each other. Therefore, the decay lengths of their edge states are not accessible.

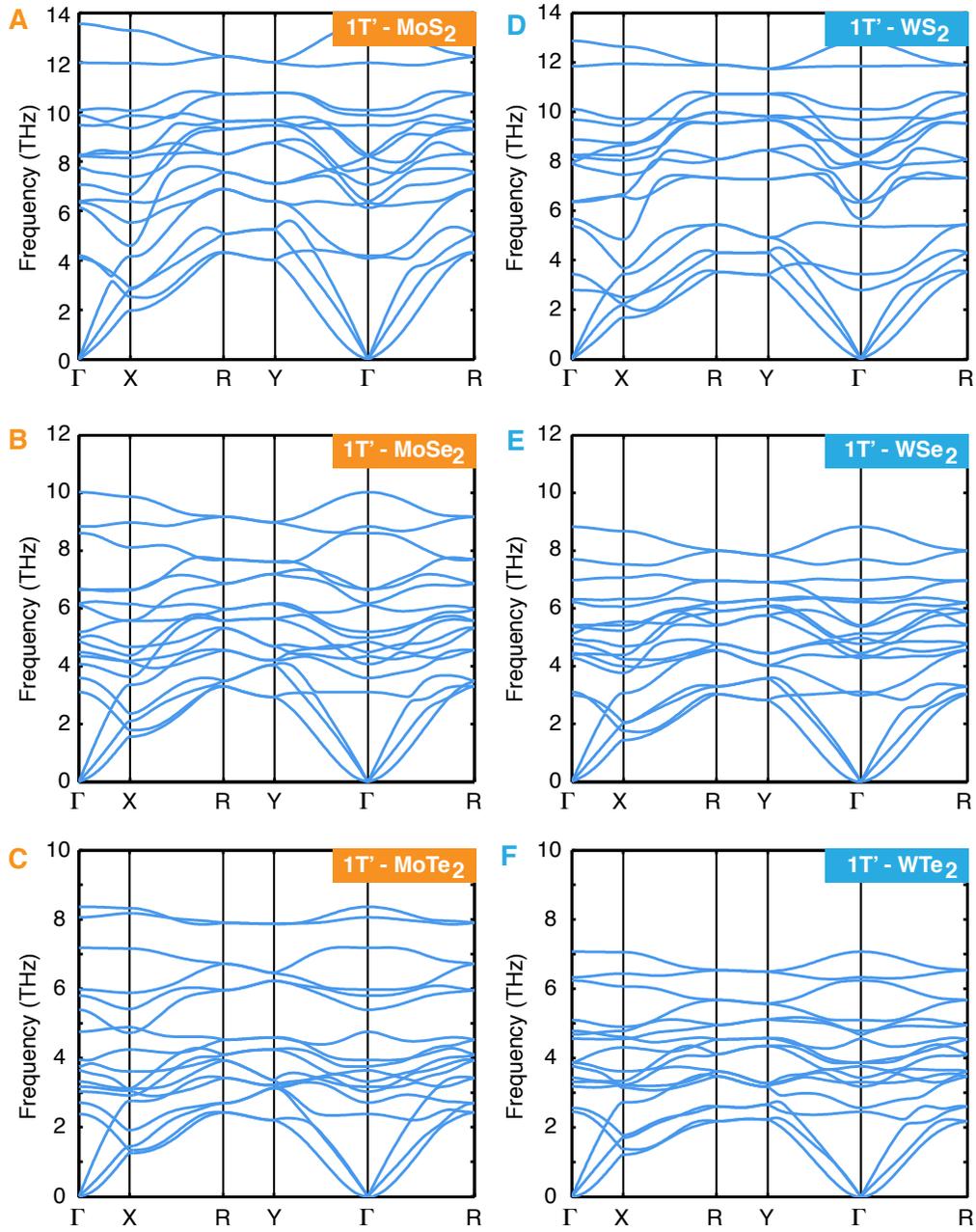

Fig. S5. Phonon dispersion curves of monolayer transition metal dichalcogenide 1T'-MX$_2$. The absence of imaginary frequency demonstrates the structural stability of all six monolayer 1T'-MX$_2$.

**References:**

1. P. Hohenberg, W. Kohn, Inhomogeneous Electron Gas. *Phys. Rev. B* **136**, B864 (1964).
2. W. Kohn, L. Sham, Self-Consistent Equations Including Exchange and Correlation Effects. *Phys. Rev.* **140**, 1133 (1965).
3. G. Kresse, J. Furthmuller, Efficiency of ab-initio total energy calculations for metals and semiconductors using a plane-wave basis set. *Comput. Mater. Sci.* **6**, 15-50 (1996).
4. G. Kresse, J. Furthmuller, Efficient iterative schemes for ab initio total-energy calculations using a plane-wave basis set. *Phys. Rev. B* **54**, 11169-11186 (1996).
5. P. E. Blöchl, Projector augmented-wave method. *Phys. Rev. B* **50**, 17953-17979 (1994).
6. J. P. Perdew, K. Burke, M. Ernzerhof, Generalized gradient approximation made simple. *Phys. Rev. Lett.* **77**, 3865-3868 (1996).
7. D. C. Langreth, M. J. Mehl, Beyond the local-density approximation in calculations of ground-state electronic-properties. *Phys. Rev. B* **28**, 1809-1834 (1983).
8. A. D. Becke, Density-functional exchange-energy approximation with correct asymptotic-behavior. *Phys. Rev. A* **38**, 3098-3100 (1988).
9. H. Monkhorst, J. Pack, Special Points for Brillouin-Zone Integrations. *Phys. Rev. B* **13**, 5188-5192 (1976).
10. A. Togo, F. Oba, I. Tanaka, First-principles calculations of the ferroelastic transition between rutile-type and $CaCl_2$-type $SiO_2$ at high pressures. *Phys. Rev. B* **78**, 134106 (2008).
11. L. Hedin, New Method for Calculating the One-Particle Green's Function with Application to the Electron-Gas Problem. *Phys. Rev.* **139**, A796-A823 (1965).
12. M. S. Hybertsen, S. G. Louie, Electron correlation in semiconductors and insulators: Band gaps and quasiparticle energies. *Phys. Rev. B* **34**, 5390-5413 (1986).
13. M. P. L. Sancho, J. M. L. Sancho, J. Rubio, Highly convergent schemes for calculation of bulk and surface Green-functions. *J. Phys. F* **15**, 851-858 (1985).
14. N. Marzari, D. Vanderbilt, Maximally localized generalized Wannier functions for composite energy bands. *Phys. Rev. B* **56**, 12847-12865 (1997).
15. X. Qian, J. Li, L. Qi, C.-Z. Wang, T.-L. Chan, Y.-X. Yao, K.-M. Ho, S. Yip, Quasiatomic orbitals for ab initio tight-binding analysis. *Phys. Rev. B* **78**, 245112 (2008).
16. L. Fu, C. L. Kane, Topological insulators with inversion symmetry. *Phys. Rev. B* **76**, 045302 (2007).
17. T. Fukui, Y. Hatsugai, Quantum Spin Hall Effect in Three Dimensional Materials: Lattice Computation of $Z_2$ Topological Invariants and Its Application to Bi and Sb. *J. Phys. Soc. Jpn.* **76**, 053702 (2007).